\documentclass{article}

\usepackage[english]{babel}

\usepackage[a4paper,top=2cm,bottom=2cm,left=3cm,right=3cm,marginparwidth=1.75cm]{geometry}

\usepackage{amssymb}
\usepackage{siunitx}
\PassOptionsToPackage{hyphens}{url}\usepackage{hyperref}
\usepackage{cleveref}
\usepackage[utf8]{inputenc}
\usepackage[right]{lineno}
\usepackage{csquotes}
\usepackage{booktabs}
\usepackage{longtable}
\usepackage{adjustbox}
\usepackage{array}
\usepackage{url}
\usepackage{titlesec}
\usepackage{authblk}
\usepackage{xcolor} % Load the xcolor package for color options

\titleformat{\subsection}
  {\mdseries\itshape\large} % Medium series, italic shape, and large font size
  {\thesubsection}{1em}{} % Numbering, spacing, and the section title itself

\usepackage[english]{babel}
\usepackage[style=apa,backend=biber,natbib=true,maxcitenames=2,uniquelist=false]{biblatex}
\DeclareNameAlias{sortname}{family-given}
\DeclareNameAlias{default}{family-given}

\renewbibmacro{in:}{}
\DeclareFieldFormat[article]{title}{\mkbibquote{#1}\addcomma}
\DeclareFieldFormat[book]{title}{\mkbibemph{#1}\addcomma}
\DeclareFieldFormat[bookinbook]{title}{\mkbibemph{#1}\addcomma}
\DeclareFieldFormat[inbook]{title}{\mkbibquote{#1}\addcomma}
\DeclareFieldFormat[incollection]{title}{\mkbibquote{#1}\addcomma}
\DeclareFieldFormat[inproceedings]{title}{\mkbibquote{#1}\addcomma}
\DeclareFieldFormat[manual]{title}{\mkbibemph{#1}\addcomma}
\DeclareFieldFormat[misc]{title}{\mkbibemph{#1}\addcomma}
\DeclareFieldFormat[thesis]{title}{\mkbibemph{#1}\addcomma}
\DeclareFieldFormat[unpublished]{title}{\mkbibquote{#1}\addcomma}
\DeclareFieldFormat[patent]{title}{\mkbibemph{#1}\addcomma}
\DeclareFieldFormat[report]{title}{\mkbibemph{#1}\addcomma}
\DeclareFieldFormat[online]{title}{\mkbibquote{#1}\addcomma}
\DeclareFieldFormat[software]{title}{\mkbibemph{#1}\addcomma}
\DeclareFieldFormat[booklet]{title}{\mkbibemph{#1}\addcomma}
\DeclareFieldFormat[periodical]{title}{\mkbibemph{#1}\addcomma}
\DeclareFieldFormat[standard]{title}{\mkbibemph{#1}\addcomma}

\DeclareFieldFormat[article]{journaltitle}{\iffieldundef{shortjournal}{\mkbibemph{#1}\addcomma}{\mkbibemph{\printfield{shortjournal}}\addcomma}}
\DeclareFieldFormat{volume}{\bibstring{volume}~#1}
\DeclareFieldFormat{number}{\bibstring{number}~#1}

\DefineBibliographyStrings{english}{
  volume = {Vol.},
  number = {No.}
}

\renewbibmacro*{volume+number+eid}{%
  \printfield{volume}%
  \setunit*{\addspace}%
  \printfield{number}%
  \setunit{\addcomma\space}%
  \printfield{eid}}

\renewbibmacro*{journal+issuetitle}{%
  \usebibmacro{journal}%
  \setunit*{\addcomma\space}%
  \usebibmacro{volume+number+eid}%
  \setunit{\addcomma\space}%
  \usebibmacro{issue+date}}

\renewbibmacro*{publisher+location+date}{%
  \printlist{publisher}%
  \iflistundef{location}
    {\setunit*{\addcomma\space}}
    {\setunit*{\addcolon\space}}%
  \printlist{location}%
  \setunit*{\addcomma\space}%
  \usebibmacro{date}}

\DeclareCiteCommand{\cite}[\mkbibparens]
  {\usebibmacro{prenote}}
  {\usebibmacro{citeindex}%
   \usebibmacro{cite}}
  {\multicitedelim}
  {\usebibmacro{postnote}}

\renewbibmacro*{cite:labelyear+extrayear}{%
  \iffieldundef{labelyear}
    {}
    {\printtext[bibhyperref]{%
       \printfield{labelyear}%
       \printfield{extrayear}}}}

\renewbibmacro*{cite:labeldate+extradate}{%
  \iffieldundef{labelyear}
    {}
    {\printtext[bibhyperref]{%
       \printfield{labelyear}%
       \printfield{extradate}}}}

\AtEveryBibitem{
  \clearfield{month}
  \clearfield{day}
  \ifentrytype{book}{
    \clearlist{location}
  }{}
}

\DefineBibliographyStrings{english}{
  andothers = {\textit{et al.},}
}

\DeclareFieldFormat[article]{volume}{\bibstring{jourvol}\addnbspace #1}
\DeclareFieldFormat[article]{number}{\bibstring{number}\addnbspace #1}
\DeclareFieldFormat[article]{volume}{Vol. #1}
\DeclareFieldFormat[article]{number}{No. #1}
\DeclareFieldFormat{url}{\bibstring{available at}\addcolon\space\url{#1}}
\DeclareFieldFormat{urldate}{\mkbibparens{accessed \addspace#1}}

\DeclareFieldFormat{urldate}{%
  \mkbibparens{accessed\space%
    \thefield{urlday}\addspace%
    \mkbibmonth{\thefield{urlmonth}}\addspace%
    \thefield{urlyear}}}

\crefformat{figure}{#2Figure~#1#3}
\Crefformat{figure}{#2Figure~#1#3}
\crefformat{table}{#2Table~#1#3}
\Crefformat{table}{#2Table~#1#3}
\crefformat{section}{#2Section~#1#3}
\Crefformat{section}{#2Section~#1#3}

\author[1$\ast$]{J.  A. Martínez-Cadena (martinezcadenajuan@gmail.com)} 
\author[1]{J.  M. Sánchez-Cerritos (jmsc@xanum.uam.mx)}
\author[2]{A.  Marin-Lopez (marinlopabi@gmail.com)}
\author[3]{M. Meraz (meraz@xanum.uam.mx)}
\author[2]{J. Alvarez-Ramirez (jjar@xanum.uam.mx)}

\affil[1]{Departamento de Matemáticas, Universidad Autónoma Metropolitana-Iztapalapa, Iztapalapa, CDMX, 09340,  México.}
\affil[2]{Departamento de Ingeniería de Procesos e Hidráulica. Universidad Autónoma Metropolitana-Iztapalapa,  Iztapalapa, CDMX, 09340, México.}
\affil[3]{Programa de Energía y Medio Ambiente y Departamento de Biotecnología. Universidad Autónoma Metropolitana-Iztapalapa, Iztapalapa, CDMX, 09340 México.}

\title{Causal wavelet analysis of ozone pollution contingencies in the Mexico City Metropolitan Area}
\date{} 

\begin{document}

\maketitle

\textbf{* Corresponding Author}.  martinezcadenajuan@gmail.com 

\begin{abstract}
In the recent two decades, the Mexico City Metropolitan Area (MCMA) has been plagued by high concentrations of air pollutants, risking the health integrity of its inhabitants. Although some policies have been undertaken, they have been insufficient to deplete high air pollutants. Environmental contingencies are commonly imposed when the ozone concentration overpasses a certain threshold, which is well above the recommended maximum by the WHO. This work used a causal version of a generalized Morlet wavelet to characterize the dynamics of daily ozone concentration in the MCMA. The results indicated that the formation of dangerous ozone concentration levels is a consequence of accumulation and incomplete dissipation effects acting over a wide range of time scales. Ozone contingencies occurred when the wavelet coefficient power is increasing, which was linked to an inti-persistence behavior. It was proposed that the wavelet methodology could be used as a further tool for signaling the potential formation of adverse ozone pollution scenarios. 
\\
\\
%add 6 keywords
\end{abstract}
%\linenumbers

\section{Introduction}
\label{sec:introduction}

The Mexico City Metropolitan Area (MCMA) has experienced huge growth in the recent five decades. The traditional urban limits have been moved outwards due to the continuous influx of immigrants from rural areas. By 2020, the number of inhabitants was more than 20 million, with a population density of about 2884 inhabitants per km2. The metropolitan urbanized area has expanded to more than 1500 km2. The combined activity of commerce, motor vehicles and industries consumes more than 50 million liters per day of combustibles, resulting in the emission of thousands of tons of air pollutants. The meteorological and topographical conditions of Mexico City are characterized by high altitude (2440 m above sea level) and tropical insolation, and the presence of primary volatile organic component (VOC) pollutants promotes the production of secondary contaminants (e.g., nitrogen dioxide and ozone), and the generation of large amounts of particulate matter. Orta-Garcia et al. (2014) warned that continuous exposure to adverse air pollution conditions is a serious threat to human health with diverse adverse consequences to economic productivity.
     A series of policies have been undertaken to control the presence of haze polluters in the air of MCMA. The use of cleaner fuels with reduced sulfur and benzene content has represented an important step toward the control of haze emissions by motor exhausts (Lopez and Mandujano, 2005). In 2014, a new vehicle traffic rule established the phasing out of 8 years and older vehicles. However, this banning was promptly reversed by the justice system in 2015. Therefore, thousands of motor vehicles were reincorporated into circulation, exacerbating car jams, slow-velocity traffic problems and automotive exhaustion emissions. The shifting of contaminant industries to outer areas has been planned for many decades, with limited outcomes. The monitoring of major pollutants has indicated that despite the pollutant concentration decrements from the 1990s, non-significant reductions have been obtained in recent years to fulfill WHO's international standards. Environmental emergencies caused by excessive levels of air pollutants have been imposed (e.g., November 12, 2022, and March 26, 2023). However, the emergencies are of short duration, one or two days, until pollutant concentrations are reduced to acceptable levels according to local criteria. 
     Molina et al. (2020) have argued that megacities, like Mexico City, have some room to manage the growing population sustainably while reducing atmospheric pollution and its impacts. To this end, appropriate planning and strong emissions control policies based on accurate scientific research should be implemented in the short term. In this line, several efforts have been devoted to achieving a close understanding of the mechanisms and interactions involved in the pollutant’s dynamics in Mexico City. Meraz et al. (2015) used rescaled range analysis for the time series of four pollutants (O3, NO2, SO2 and PM10) to find the presence of long-range memory effects, from days to months. Cardenas-Moreno et al. (2020) studied the dynamics of particulate matter, finding a twofold power law behavior in the power spectrum of all the series, indicating the co-existence of two different mechanisms underlying the time dynamics of PM10. Bing et al. (2015) showed that the ozone dynamics contain certain predictability and elaborated a scheme for forecasting based on artificial intelligence. Aguilar-Velázquez and Reyes-Ramírez (2018) used wavelet analysis to study multiday extreme ozone episodes for 2015-2016. It was found that multiday episodes exhibit periods higher than four days. Besides, it was found that for such periods, NO2, CO and O3 were correlated in a multi-temporal clustered form. Ramos-Ibarra and Silva (2020) used the Hodrick-Prescott filter with a Kalman filter to forecast air pollutants in Mexico City. It was found that, except CO and NO2, the trend of the remaining pollutants was far from permissible limits. Rios et al. (2023) used wavelet analysis to assess the impact of biomass burning on the air quality in Mexico City.
     In the recent decade, the MCMA has been subjected to a series of ozone environmental emergencies, which mandate the restriction of automobile circulation. However, these contingencies are imposed when the air pollution is at dangerous levels while lasting one or two days at most. An interesting question is to assess whether the evolution of the ozone concentration provides signals to warning the occurrence of adverse atmospheric conditions, and in this way to take actions oriented to reduce adverse effects in socioeconomical activity and human health. This work aims to use a causal wavelet analysis on the ozone concentration dynamics to gain insights into the possible warnings concerning adverse pollution conditions. The task is to show that causal wavelet analysis is a suitable framework to characterize the dynamics of ozone conditions and to link the results with the potential occurrence of atmospheric pollution emergencies.

\section{Data}
\label{sec:data}
The MCMA is located at 19.4°N, 99.1°W, and 2240 m asl. The region is surrounded by mountain ridges exceeding 5000 m asl. The topography is complex (Figure 1.a), which magnifies the impact of thermal inversions that inhibit winds while favoring intense air pollution (Carreon-Sierra et al., 2015). 
Figure 1.b illustrates a typical day in MCMA with high ozone concentration levels. Air pollutants are monitored by the Red Automática de Monitoreo Ambiental (RAMA) through 28 stations. The data on ozone pollution was obtained from the publicly accessible site  \href{www.aire.cdmx.gob.mx/}{www.aire.cdmx.gob.mx/}(4899 daily-averaged observations). Measurement gaps were filled with the method reported by Diosdado et al. (2013). The fraction of gaps was not higher than 0.5\%, and the number of consecutive gaps was not higher than 12 registers. Environmental contingencies are extreme events declared when the ozone concentration exceeds the 150 IMECAS threshold. The occurrence of the EC triggered by ozone was obtained from the official report located on  \href{www.aire.cdmx.gob.mx/}{www.aire.cdmx.gob.mx/}.

\begin{figure}[ht]
 \centering
 \makebox[\textwidth][c]{\includegraphics[width=.6\textwidth]{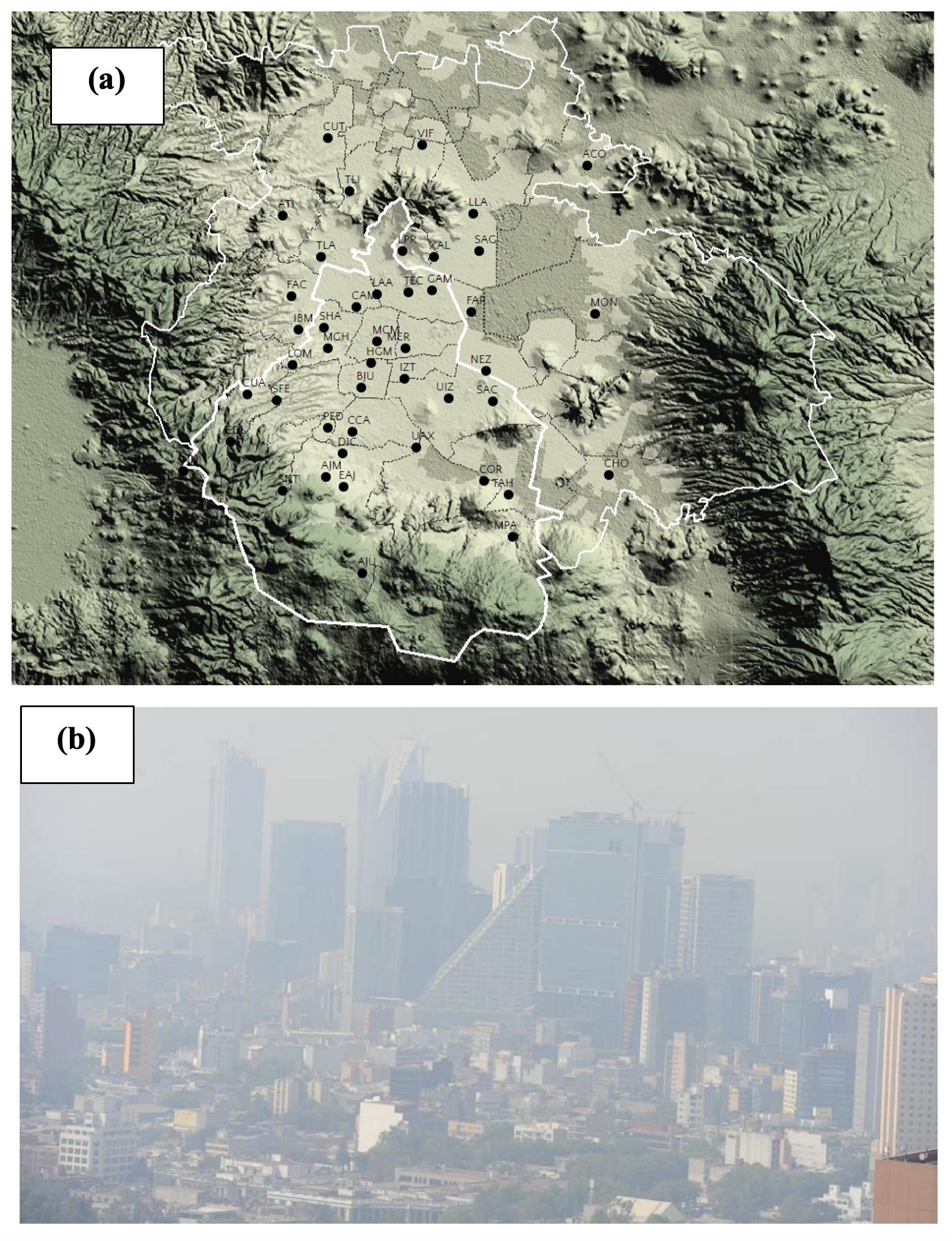}}%
 \caption{(a) Map of Mexico City and the surrounding areas. The orange rectangle denotes the approximate location of the Mexico City center region. (b) Typical scene of a day in CDMX with high ozone concentration and environmental contingency.}
 \label{stations}
\end{figure}

\section{Methodology}
\label{sec:methodology}
The methodology in this study is based on wavelet analysis. Wavelets provide information about the evolution of a signal across different frequency and time scales. A complex-valued wavelet $\psi(\sigma;u,s)$ is a square integral function with a scale $s$, location $u$ and time $t$, which can be defined as a function $\frac{1}{\sqrt{s}}\psi(\sigma;u,s)$, where the variable 
\begin{equation}
\sigma = \frac{u-t}{s}
\end{equation}
can be seen as a dimensionless time scale. For a given time function $x(t)$, the continuous wavelet transform is obtained by the projection of the function $x(t)$ on the conjugate $\psi^*(\sigma;u,s)$ of $\psi(\sigma;u,s)$ as follows:

\begin{equation}\label{TW}
W_x(u,s)=\frac{1}{\sqrt{s}} \int_{-\infty}^{+\infty} \psi^*\left( \frac{u-t}{s} \right) x(t) dt
\end{equation}

The continuous wavelet transform decomposes the time function $x(t)$ into scale components, which can be used to reconstruct the original function as follows:

\begin{equation}
x(t) = \frac{\int_{0}^{+\infty}\int_{-\infty}^{+\infty  } W_x(u,s) \psi^*\left( \frac{u-t}{s}; u,s \right) du ds }{s^2 C_\psi}
\end{equation}

A considerable number of wavelet functions have been proposed in recent decades (Guo et al., 2022). Here, we will consider wavelet functions of the form

\begin{equation}
\psi(\sigma) = M(\sigma)e^{i\sigma}
\end{equation}

The function  $\psi(\sigma)$ is formed by an oscillatory pattern $e^{i\sigma}$ modulated by the function $M(\sigma)$.  The above wavelet structure has the following attractive features. From the Euler identity, one can write the following expression: 

\begin{equation}
\psi(\sigma) = M(\sigma) (\cos (\sigma) + i \sin (\sigma)).
\end{equation}

In terms of the scale s and the time location $u$, one has

\begin{equation}
\psi(\sigma; u,s) = M\left( \frac{u-t}{s} \right) \left( \cos \left( \frac{u-t}{s} \right) + i \sin \left( \frac{u-t}{s} \right) \right).
\end{equation}

For a real-valued modulation function, the wavelet coefficient is given as follows:

\begin{equation}
W_x(u,s)=W_{x,Re}(u,s) + iW_{x,Im}(u,s)=
\end{equation}
$$
=\frac{1}{\sqrt{s}} \int_{-\infty}^{+\infty  }  M\left( \frac{u-t}{s} \right) \cos \left( \frac{u-t}{s} \right) x(t) dt - i\frac{1}{\sqrt{s}} \int_{-\infty}^{+\infty  }  M\left( \frac{u-t}{s} \right) \sin \left( \frac{u-t}{s} \right) x(t) dt
$$

The wavelet structure given by Eq. (6) is interesting since Eq. (7) resembles the computation of the coefficients in classical Fourier analysis with $\omega=s^{-1}$. For a unit modulation function (i.e., $M(\sigma)=1$), $W_x (u,s)$ corresponds to the complex Fourier coefficient for the frequency $\omega$.  In this way, the function $M(\sigma)$ weights the contribution of the Fourier basis functions $\cos(t)$ and $\sin(t)$ for the time $t$. Note that $\frac{1}{\sqrt{s}}$ acts as a normalization factor to account for the “support” of the wavelet $M(\sigma)$. The Fourier analogy allows us to interpret the wavelet coefficient in the following form. The magnitude,  also called power, $|W_x (u,s)|$ reflects the relative contribution of the components at the scale $s$ and the time location $u$. On the other hand, the phase 

\begin{equation}
\theta_x (u,s) = \tan^{-1} \left(   \frac{W_{x,Re}(u,s)}{W_{x,Im}(u,s)} \right) 
\end{equation}

quantifies the contribution of the even components relative to the odd components of the times function $x(t)$. For an even (resp., odd) function, one has that  $W_{x,Im}(u,s)=0$ (resp, $W_{x,Re}(u,s)=0$), such that $|\theta_x (u,s)|\to \frac{\pi}{2}$ (resp, $|\theta_x (u,s)|\to 0$). Even and odd functions are reflected by the wavelet components $\cos (\sigma t)$ and $\sin (\sigma t)$.  For $|\theta_x (u,s)|\to 0$, whenever the value $x(t)$ has been up (resp, down), it is more likely that it will be up (resp, down) in the close past or future. In contrast, if $|\theta_x (u,s)|\to \frac{\pi}{2}$, one has the opposite pattern in the sense that whenever the value $x(t)$ has been up (resp, down), it is more likely that it will be down (resp, up) in the close past or future. In this regard,  $|\theta_x (u,s)|\to 0$ and $|\theta_x (u,s)|\to \frac{\pi}{2}$ reflect anti-persistence and persistence, respectively.

     The specification of the modulation function $M(\sigma)$ is a key issue of the performance of the wavelet analysis. The modulation function $M(\sigma)$ determines the degree of locality of the wavelet analysis. If $M(\sigma)=1$ for all $t$, then the coefficient $W_x (u,s)$ reflects the pattern of the oscillatory behavior averaged over the whole-time domain (i.e., classical Fourier analysis). To have a local analysis, one should have that a) $M(0)=1$ and b) $M(\sigma)$ is a decaying function of $ |\sigma|$. That is, times far from the time location u should have a minor (decreasing) contribution to the wavelet coefficient $W_x (u,s)$. We consider the one-parameter Mittag-Leffler function as a modulator of the oscillatory behavior

\begin{equation}
M(\sigma)= E_\alpha (-|\sigma|^\alpha) = \sum_{k=0}^\infty \frac{-|\sigma|^ {k\alpha}}{ \Gamma(1+k\alpha)}
\end{equation}

where $\Gamma(.)$ is the Gamma function. In the limit as $\alpha \to 1$ one recovers the Taylor series expansion of the standard exponential function. The numerical evaluation of the Mittag-Leffler function can present some troubles, especially for large values of the index k where the gamma function can exhibit overflow stacks. The stretched exponential function 

\begin{equation}
 E_\alpha (-|\sigma|^\alpha)  \approx e^{-|\sigma|^\alpha}
\end{equation}

is an acceptable approximation for relatively large values of $\sigma$. The case $\alpha = 2$ gives $M(\sigma)=e^{-|\sigma|^\alpha}$,  which is a Gaussian function, and the resulting modulator leads to the Morlet wavelet expression. In this way, one has that

\begin{equation}
\psi(\sigma; u,s) =  E_\alpha (-|\sigma|^2) e^{i \sigma}
\end{equation}

can be seen as a generalization of the Morlet wavelet. The Mittag-Leffler function has been used a kernel for fractional integral and derivative (Atangana and Baleanu, 2016). Hence, the wavelet $\psi(\sigma; u,s)$ can be also used to reflect the fractional structure of a given time series.

\subsection{Causal wavelet analysis}

The Mittag-Leffler modulation function $E_\alpha (-|\sigma|^\alpha)$ is a symmetric function, which means that for a given location past and future values have equal contribution to the wavelet coefficient $W_x (u,s)$. This would imply that the behavior in future times impacts the time series behavior in past times. However, such a modulation structure violates the causality of the time series. In this regard, the modulation function should be modified to consider only past times in the computation of the wavelet coefficient. Following Szu et al. (1992), the causality modification can be achieved by taking the modulation function only for past times:

\begin{equation}
M(\sigma)= \left\{ \begin{array}{lcc}
E_\alpha (-|\sigma|^\alpha) & \mbox{for} & \sigma \geq 0 \\ \\ 
0 &  \mbox{for} & \sigma \leq 0 \end{array} \right.
\end{equation}

By doing so, the wavelet coefficient is given by

\begin{equation}
W_x(u,s) = \frac{1}{\sqrt{s}}  \int_{-\infty}^{u}    \psi^*  \left( \frac{u-t}{s}\right)  x(t)dt
\end{equation}

In this way, the causal wavelet function  $ \psi^* (\sigma)$ plays the role of a fading memory kernel acting on the time series $x(t)$. Past values of $x(t)$, $t<u$, have a fading contribution to the integral in Eq. (13). The sharper the decaying of the modulation, the more local nature of the wavelet coefficient $W_x (u,s)$. On the other hand, the term $\frac{1}{\sqrt{s}}$ can be interpreted as a factor that accounts for the time scale effect in the integration.

     The exponent $\alpha$ affects the sharpness of the decaying modulation function $M(\sigma)$. Figure 2.a shows the behavior of the Mittag-Leffler modulation function for five different values of the exponent $\alpha$. As already mentioned above, the modulation function approaches $M(\sigma)=1$ as the exponent  $\alpha \to 0$.  For small values of  $\alpha$, the locality of the wavelet analysis is decreased. In contrast, the modulation function becomes a sharper decaying function for large values of  $\alpha$, which in turn strengthens the locality of the wavelet function. The corresponding even (i.e., real) and odd (i.e., imaginary) parts of the wavelet function are illustrated in figures 2.b and 2.c, respectively. If one uses the stretched exponential approximation (10), a causal version of the complex Morlet wavelet function is recovered for $\alpha=2$.

\begin{figure}[ht]
 \centering
 \makebox[\textwidth][c]{\includegraphics[width=.6\textwidth]{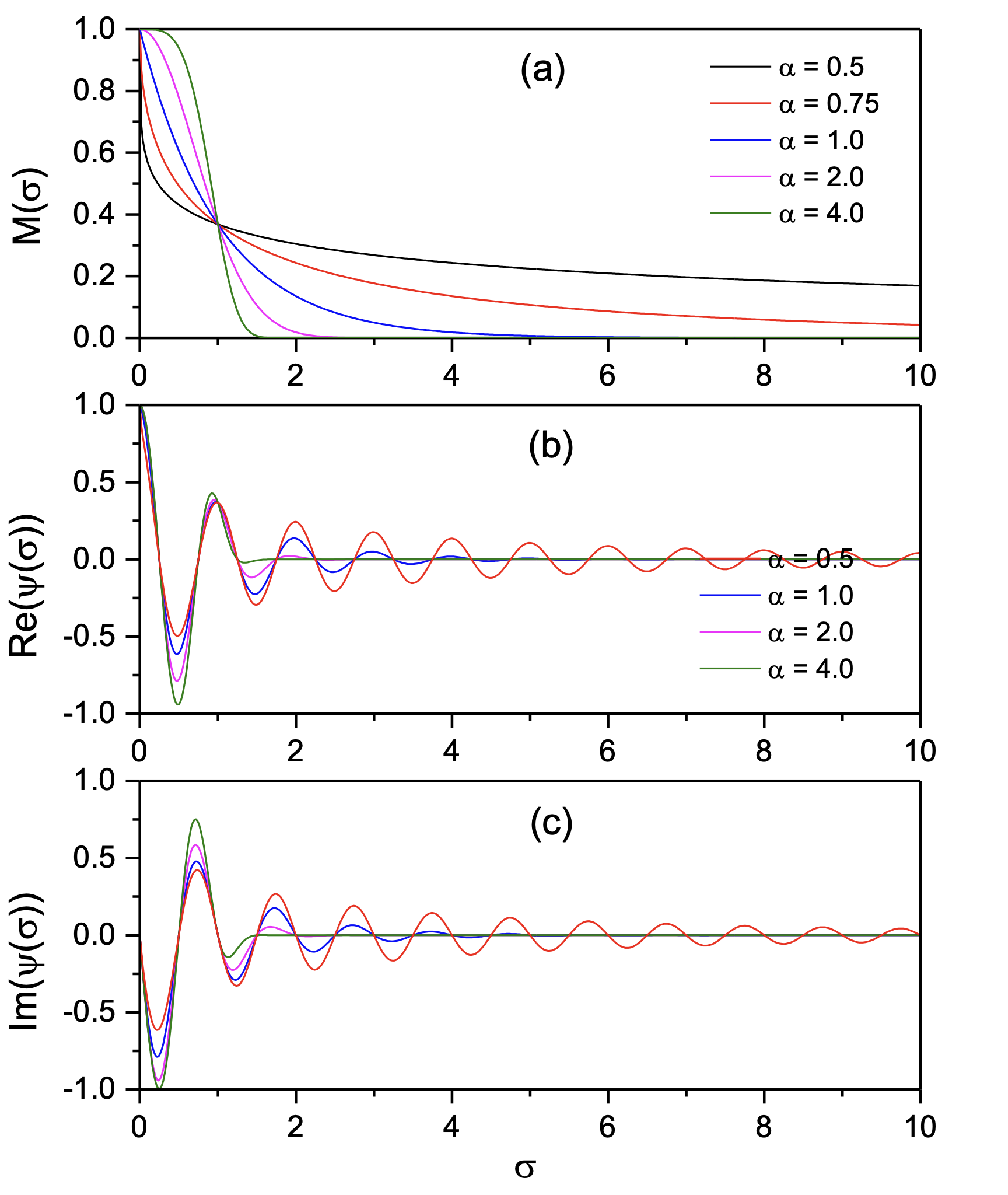}}%
 \caption{(a) Effect of the exponent with the parameter $\alpha$ in the behavior of the Mittag-Leffler modulation function. The even and odd components of the wavelet function for different values of the parameter $\alpha$ are shown in panels (b) and (c), respectively.}
 \label{stations}
\end{figure}

\section{Results and discussion}
     
The ozone time series presents positive skewness (0.57), indicating a trend to extremely high values of the pollutants rather than a trend showing reduced values. Fat tails are also contained in the ozone distribution, as shown in terms of kurtosis (0.11) and Shapiro-Wilk statistics (0.98). Figure 3 exhibits the ozone time series in the period from January 2010 to June 2023. The maximum and minimum of the ozone cycle are found for the winter-spring and summer-autumn seasons, respectively. 
     
Tropospheric ozone is found at the surface level, in urban areas it is produced when nitrogen oxides (NOX) and volatile organic compounds (VOCs) react in the atmosphere in the presence of sunlight. Ozone is a strong oxidant that in high concentrations irritates the eyes and respiratory tract, reducing respiratory function (Bell et al., 2006). The mean ozone concentration for the period 2010-2015 was 48.20$\pm$18.91 µg/m3. However, the period from 2016-2023 exhibited an increase to 53.21$\pm$20.21 µg/m3, which was caused by the end of the partial banning of circulation of vehicles older than 8 years. This action increased the motor vehicle number by about 25\%, which was reflected by an increase in the peak ozone concentration in the winter-spring periods. The Covid-19 lockdown in the first semester of 2020 did not have a marked impact on the ozone concentration. In this period, the ozone concentration trend remained the same as in the last five pre-pandemic years. Hernandez-Paniagua et al. (2021) used truncated Fourier analysis to find that the ozone concentration did not decrease ($p<0.05$) from the baseline at any site despite the total Covid 19 lockdown. 
     
\subsection{Wavelet analysis}
     
The vertical lines in Figure 3 denote the occurrence of environmental contingencies triggered by high ozone concentrations. Most environmental contingencies occurred when the ozone concentration was at the highest values of the winter-spring season. Year 2020 was an exception since the only contingency was declared in August when the ozone concentration did not achieve the highest values. Although the 2020 winter-spring season witnessed ozone concentration values quite beyond the WHO limit, no contingency was declared maybe because it was not relevant since most Mexico City inhabitants were under lockdown due to the Covid-19 disease. The data shown in Figure 3 indicates that environmental contingencies are likely to be present in every winter-spring season. An interesting issue is to explore whether the oscillatory behavior of the ozone concentration hides some valuable insights into the occurrence of ozone emergencies. Information in this line would provide guidelines to technicians for implementing policies and pre-contingency scenarios before the ozone pollution achieves emergency conditions. In the following, the causal wavelet analysis will be used to address the abovementioned issue. 

\begin{figure}[ht]
 \centering
 \makebox[\textwidth][c]{\includegraphics[width=.6\textwidth]{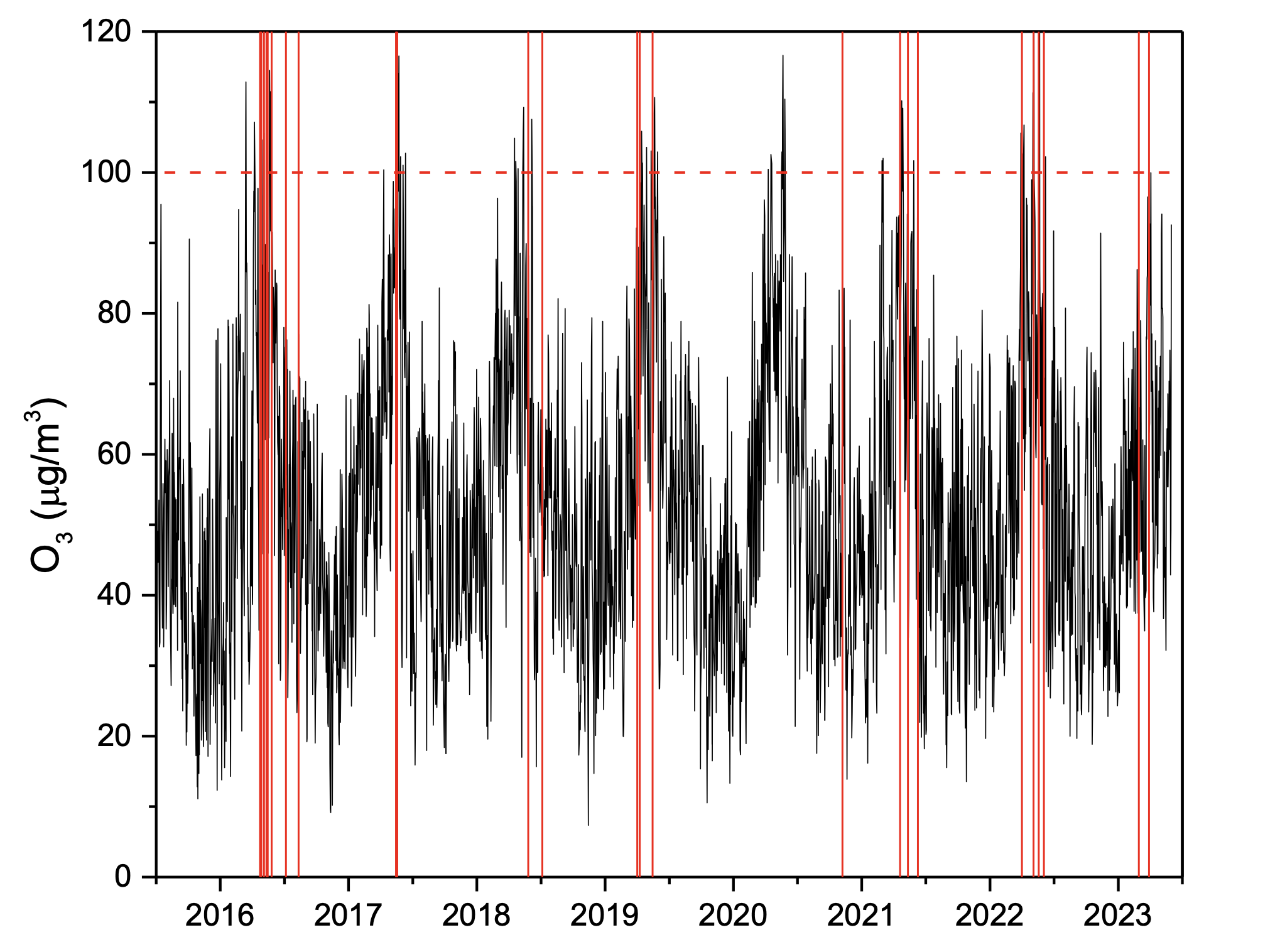}}%
 \caption{Average daily variations of ozone pollutant in Mexico City.  The horizontal dotted red line denotes the maximum daily exposure according to the 2021 WHO guidelines. The vertical lines denote the environmental contingencies triggered by high ozone concentration.}
 \label{stations}
\end{figure}

     Figure 4.a presents the power scalograms for the non-causal wavelets computed with $\alpha$=2.0. The high-power regions exhibit a symmetry-like pattern concerning the time location. This pattern is caused by the bidirectional computation of the wavelet component that accounts for both past and future dynamics. In contrast, the power scalogram computed with the causal wavelet shows patterns that are biased toward the future (Figure 4.b). The wavelet coefficient computation is based only on past information, which is projected to future times for high scales. In this way, an event occurring at a given time does not impact the past, but rather is propagated to the future along a wide range of time horizons. The above results indicated that causal wavelet analysis is the proper tool to characterize the multiscale behavior of time series dynamics. Figure 5 presents the power scalograms for four different values of the exponent $\alpha$. The increase of the exponent provides a more local wavelet function (Figure 2), which is reflected in the scalogram as more resolved regions of high-power values. The scalogram with $\alpha$=0.5 exhibits a band in the scale region of about 50-70 days. By increasing the value of $\alpha$, one obtains scalograms with increased localness where the otherwise band at 50-70 days is resolved as individual bands emerging at relatively low scales and being propagated in time to higher scales. The differences between the cases $\alpha$=2.0 and $\alpha$=4.0 are minimal, which suggests that a value of $\alpha$=2.0 can scalograms with well-resolved local information.

\begin{figure}[ht]
 \centering
 \makebox[\textwidth][c]{\includegraphics[width=.7\textwidth]{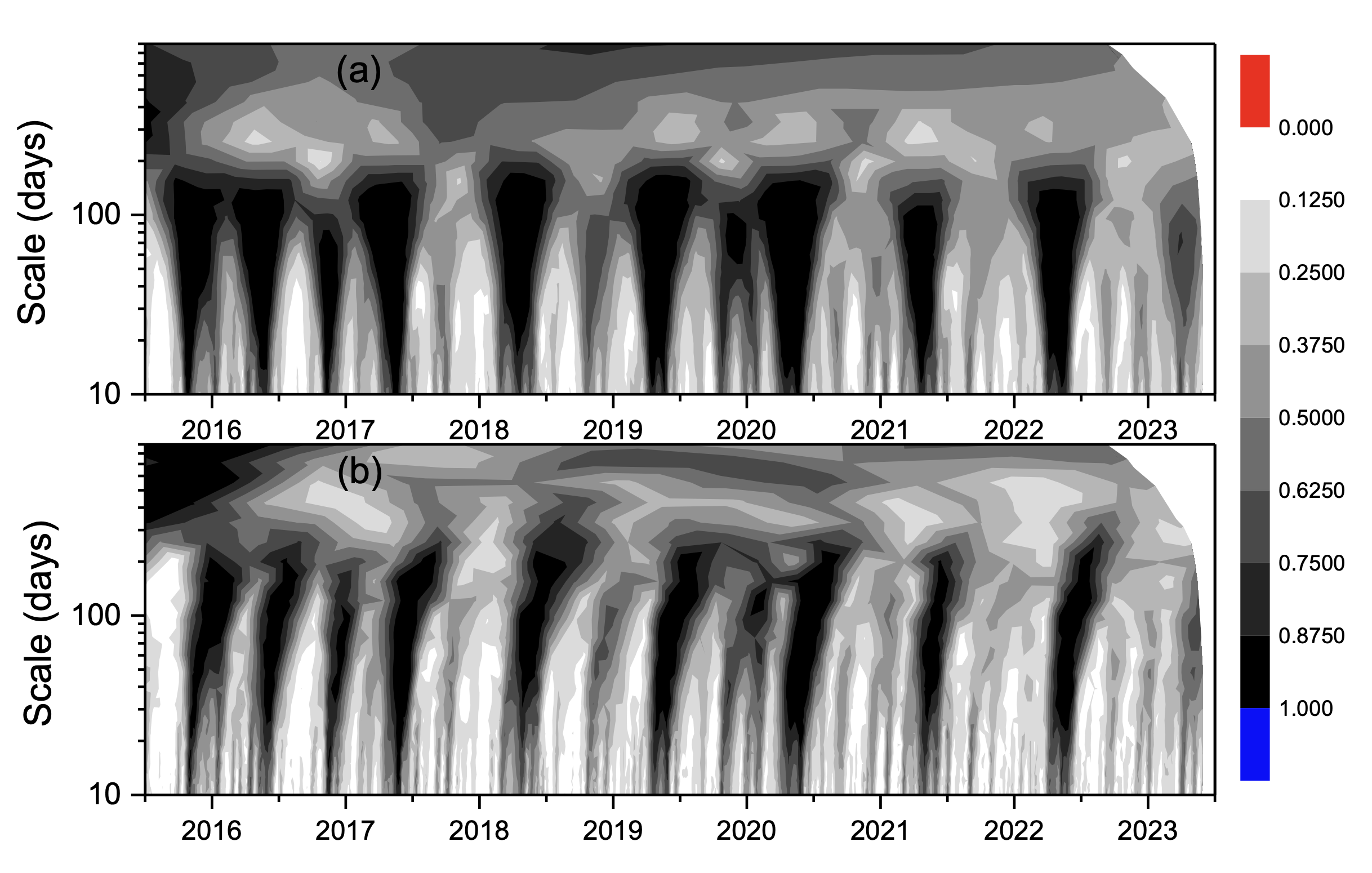}}%
 \caption{Power scalograms obtained with $\alpha$=2.0. (a) Bidirectional (i.e., noncausal) and (b) causal wavelets. }
 \label{stations}
\end{figure}

\begin{figure}[ht]
 \centering
 \makebox[\textwidth][c]{\includegraphics[width=.7\textwidth]{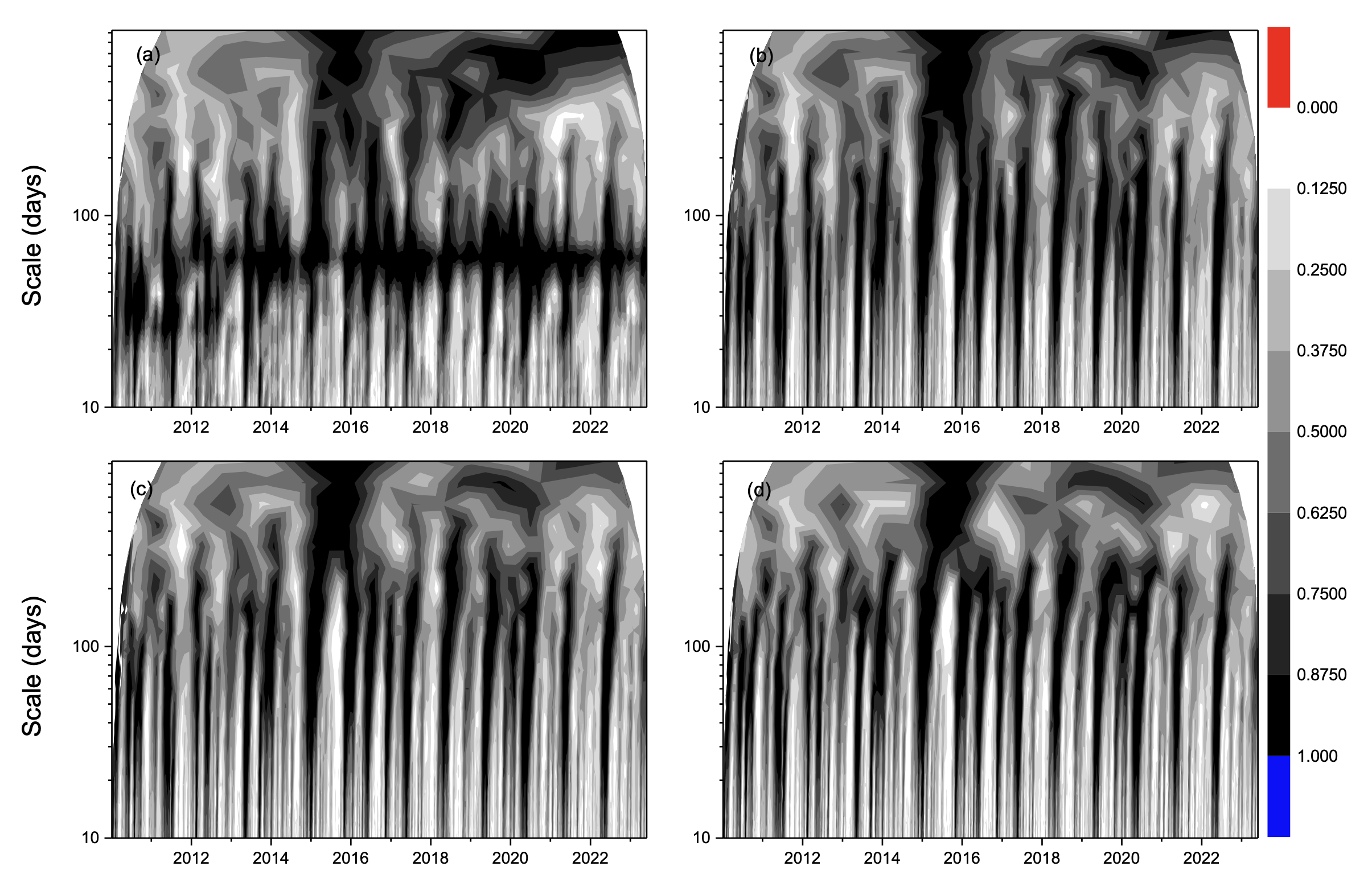}}%
 \caption{Scalogram of the ozone concentration for four different values of the exponent $\alpha$: (a) 0.5, (b) = 1.0, (c) 2.0 and (d) = 4.0.}
 \label{stations}
\end{figure}

Figure 6.a compares the causal scalogram pattern with the occurrence of the environmental contingences. Except for the emergencies of 2015 and 2020, the ozone emergencies occurred in high-power vertical bands. The emergencies in 2016, 2017, 2019, 2021 and 2022 occurred when the magnitude of the power coefficient at relatively small scales increased to relative values of about 0.85-1.0. The impact of these emergencies propagated through a wide range of scales up to about 250 days. The largest effect propagation occurred in the last months of 2014, when the pattern propagated to scales of about 900 days and persisted for the next year. The pattern exhibited in Figure 6.a suggests that the presence of high ozone concentrations is not the consequence of local-in-time events, but rather the reflection of the accumulation and lack of sufficient dissipation mechanisms along a relatively wide time horizon. Figures 7.a and 7.b show respectively the variation of the wavelet power for scales of 10 and 40 days. The agreement of the ozone contingencies and the power peaks can be observed. In some cases, the impact of the ozone is so severe in a year that is propagated to the next year via large-scale effects (e.g., 2019-2020). Interestingly, no ozone contingencies were declared in 2020 despite the high ozone concentration and it reflect high-power values. The reason may be that it was not necessary because most of the CDMX inhabitants were under lockdown due to the Covid-19 pandemic.
     Figure 6.b displays the scalogram of the phase absolute value. One has that $|\theta_x (u,s)|$ can be seen as an index of anti-persistence. Bands of low and high values of $|\theta_x (u,s)|$ alternate in time. The ozone contingencies occurred in general when the phase was close to $\frac{\pi}{2}$ over a wide range of scales. This suggests that an ozone crisis is linked to highly oscillatory patterns when the concentration achieves high values. Low values of $|\theta_x (u,s)|$ appeared in the phase previous to the occurrence of an ozone emergency, reflecting a persistency pattern when increasing concentration is likely to be followed by increasing concentration. In this regard, the phase $|\theta_x (u,s)|$ can be seen as a warning of the advent of an ozone contingency.

\begin{figure}[ht]
 \centering
 \makebox[\textwidth][c]{\includegraphics[width=.7\textwidth]{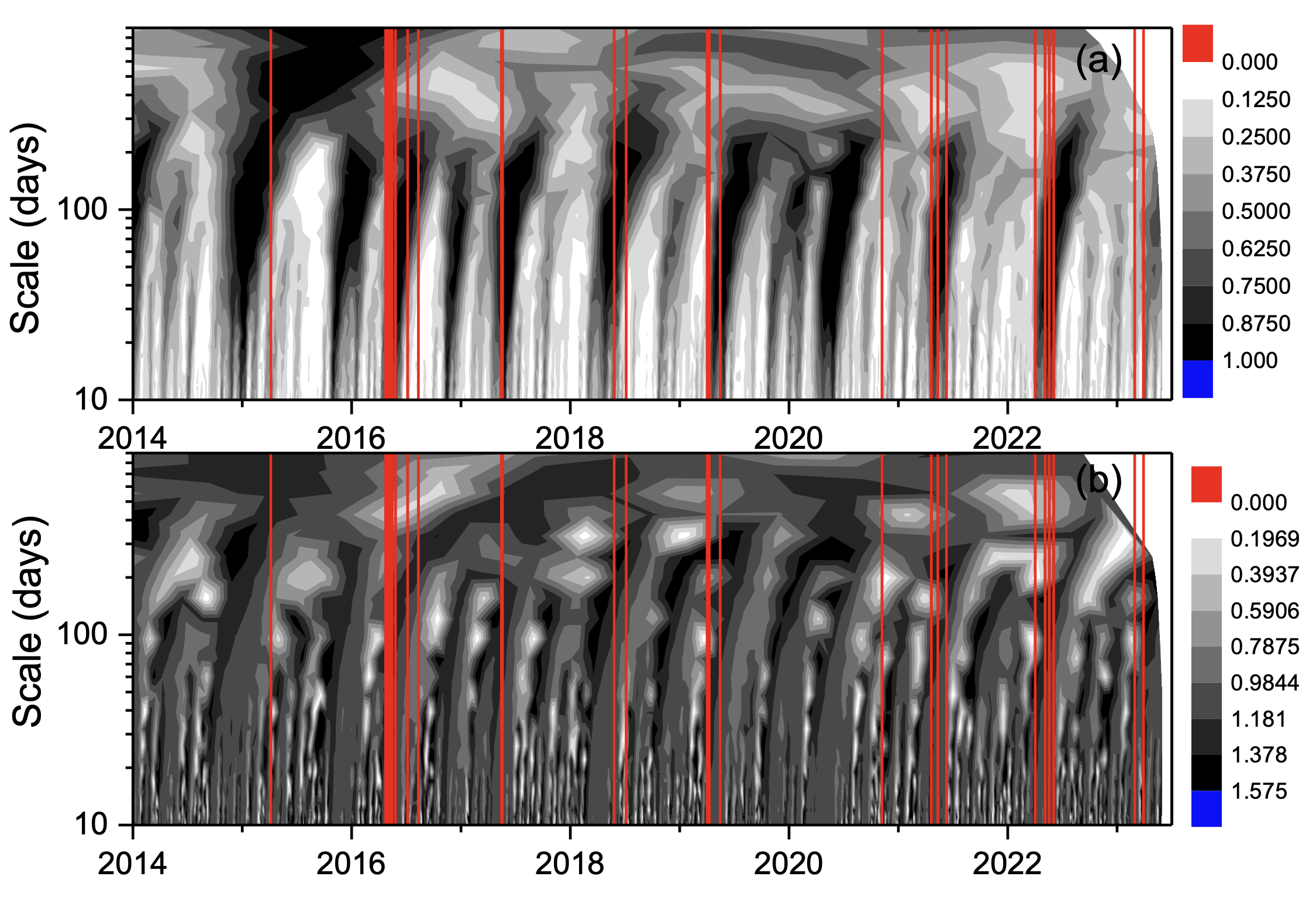}}%
 \caption{(a) Power, and (b) phase scalograms obtained with the causal wavelet for $\alpha$=2.0. The vertical lines denote the occurrence of environmental emergencies triggered by high ozone concentrations. }
 \label{stations}
\end{figure}

\begin{figure}[ht]
 \centering
 \makebox[\textwidth][c]{\includegraphics[width=.6\textwidth]{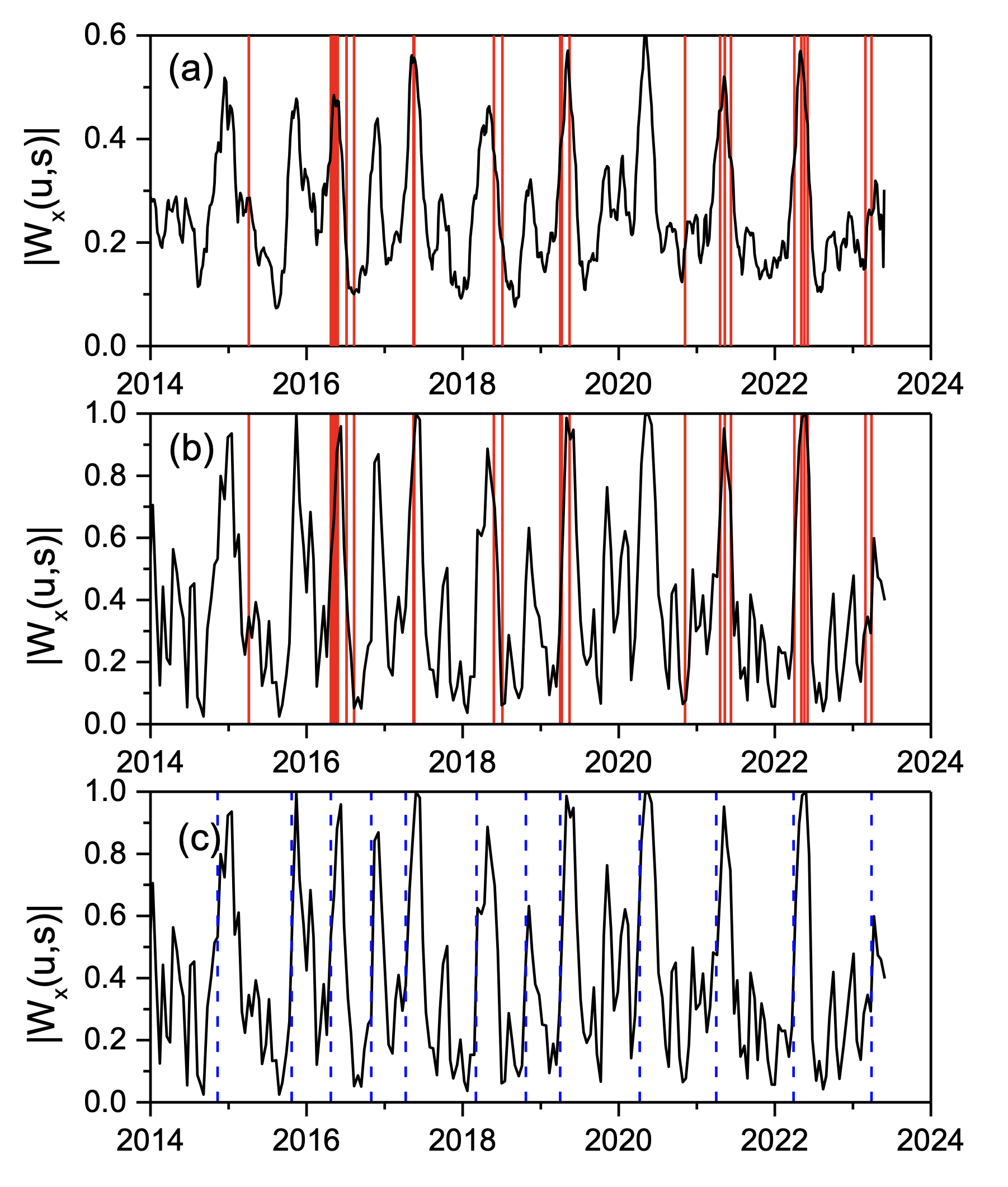}}%
 \caption{Wavelet power as function of the time for scales of (a) 10 and (b) 30 days. The vertical lines denote the occurrence of environmental emergencies triggered by high ozone concentrations. (c) Same as in panel (b), but the vertical lines signal the formation of adverse ozone pollution scenarios.}
 \label{stations}
\end{figure}
     
\subsection{Discussion}

The declaration of an environmental contingency is an undesirable event since, besides recognizing that pollutants have reached dangerous concentration levels, it negatively impacts the socio-economic activity of the metropolitan region. In this regard, policymakers can exhibit an aversion to imposing restrictions on automobile circulation. It is not surprising at all that despite ozone concentrations being high, and the wavelet power and phase results indicating that emergency conditions are being formed, environmental contingencies are not declared. This is the case, for instance, in the last months of 2015 and 2016 when ozone concentration and wavelet analysis signaled emergency conditions. A key feature of contingencies is that their declaration is made when the pollution conditions are dangerous for human health. The wavelet analysis revealed that such adverse conditions were built up over several days and along different time scales. The wavelet analysis presented above can be used as an additional tool for pre-signaling the occurrence of adverse pollution conditions. An ozone contingency signal can be taken when the wavelet power reach a certain value. For instance, Figure 7.c illustrates the possible contingency signaling for the power variation of the scale of 40 days. In this case, a warning flag was erected when the power trespassed the 0.5 value. Under this scenario, one would have 14 contingency warnings in the last 10 years. In principle, policies like this one would reduce the occurrence of ozone emergencies by designing proper environmental policies (e.g., fossil fuel usage, restricted automobile usage, etc.). Chavez-Baeza and Sheinbaum-Pardo (2014) postulated two scenarios for pollutant mitigation: (a) increasing fuel efficiencies and introducing new technologies for vehicle emission controls, and (b) modal shift from private car trips to a bus rapid transport system. Fuel efficiencies and better controls for vehicle emissions have been implemented in the recent few years. Also, preventive actions to reduce the impact of seasonal wildfires have been undertaken. On the other hand, the shift to a bus rapid transport system has shown important advances via the installation of, e.g., exclusive lanes for bus transport and an increase of the metro lines. Such public policy actions have had only a marginal impact as revealed by the persistent levels of health-threaten polluters and the undisrupted mutual correlation between contaminants (Davis, 2017). Although the Covid-19 lockdown led to a reduction in emissions, the effect was of short duration, and the pollution patterns returned rapidly to the pre-pandemic levels. Proper handling of the air pollution problems in Mexico City should involve tight controls (Goddard, 1997), such as more extract norms for vehicle mobility (Davis, 2008), reduction of fossil fuel utilization and a huge increase in the number and efficiency of the public transport system. In this line, Bel and Holst (2018) found statistical evidence that Mexico City's bus rapid transport (BRT) network leads to reducing emissions of CO, NOX, and PM10. However, the impact of the BRT may be insufficient in the medium and long run given the overthrow of pollutant emissions by inefficient private cars and massive transportation systems.

\section{Conclusion}

This work used causal wavelet analysis to characterize the dynamical behavior of ozone pollution in the Mexico City Metropolitan area. A complex generalization of the Morlet wavelet via the Mittag-Leffler function was used to compute the power and phase scalograms. It was revealed that wavelet analysis can provide signals before the occurrence of an ozone pollution contingency. Both power and phase scalograms showed scale patterns linked to the advent of ozone concentration persisting for several weeks. The analysis reported in the present study should be considered as a further step toward the design of policies oriented to mitigating the adverse effects of ozone pollution on human health.

\vspace{1cm}

\textbf{Authors declared no conflict of interest} 

\medskip

\textbf{We have not received any funding for this research} 

\medskip

\textbf{Generative AI tools were not used for the writing of the manuscript}

\printbibliography

\end{document}